\documentclass[12pt]{article}

\usepackage[margin=1in]{geometry}
\usepackage{setspace}
\usepackage{booktabs,array,ragged2e,tabularx}
\newcolumntype{L}[1]{>{\RaggedRight\arraybackslash}p{#1}}
\usepackage[skip=6pt, labelfont=bf, textfont=it, font=normalsize]{caption}
\usepackage{pdflscape}

\usepackage[hidelinks]{hyperref}
\usepackage{xurl}
\usepackage{csquotes}
\usepackage[style=apa, backend=biber, sortcites=true]{biblatex}
\DeclareLanguageMapping{english}{english-apa}
\usepackage{titling,authblk}
\pretitle{\centering\bfseries\large}
\posttitle{\par\vspace{-0.5em}}

\usepackage{authblk}
\title{AI Generated Child Sexual Abuse Material---What’s the Harm?}
\author[1]{Caoilte Ó Ciardha}
\author[2]{John Buckley}
\author[3]{Rebecca S. Portnoff\thanks{Dr Portnoff’s contribution to this paper is made in a personal capacity and does not necessarily reflect the views or positions of Thorn or its partners.}}
\affil[1]{Senior Research Fellow, University of Kent, UK}
\affil[2]{Digital Child Safety Expert}
\affil[3]{Vice President of Data Science, Thorn}
\date{}  % no date

\begin{document}
\maketitle
\vspace{-2.5\baselineskip} % tweak as needed
\begin{abstract}
%% Text of abstract
The development of generative artificial intelligence (AI) tools capable of producing wholly or partially synthetic child sexual abuse material (AI CSAM) presents profound challenges for child protection, law enforcement, and societal responses to child exploitation. While some argue that the harmfulness of AI CSAM differs fundamentally from other CSAM due to a perceived absence of direct victimization, this perspective fails to account for the range of risks associated with its production and consumption. AI has been implicated in the creation of synthetic CSAM of children who have not previously been abused, the revictimization of known survivors of abuse, the facilitation of grooming, coercion and sexual extortion, and the normalization of child sexual exploitation. Additionally, AI CSAM may serve as a new or enhanced pathway into offending by lowering barriers to engagement, desensitizing users to progressively extreme content, and undermining protective factors for individuals with a sexual interest in children. This paper provides a primer on some key technologies, critically examines the harms associated with AI CSAM, and cautions against claims that it may function as a harm reduction tool, emphasizing how some appeals to harmlessness obscure its real risks and may contribute to inertia in ecosystem responses.
\end{abstract}

%% Keywords

%% main text

\section{Introduction}
\label{sec1}

The emergence of generative artificial intelligence child sexual abuse material (AI CSAM) has been driven by the growing accessibility and sophistication of AI technologies capable of creating increasingly realistic synthetic images and videos. While the misuse of generative AI tools for explicit content began with earlier technologies (e.g., face swapping tools), the introduction of open-source diffusion models in 2021–2022 marked a significant step-change, enabling wider misuse by individuals with and without technical expertise \parencite{europol2025a}. Reports emerging in 2023 highlighted how vulnerabilities in generative AI tools, including potential bypasses of safety mechanisms during use, were contributing to the proliferation of AI CSAM (Internet Watch Foundation [IWF], \citeyear{foundation2023a}; Thiel et al., \citeyear{thiel2023a}). For instance, the IWF identified over 20,000 suspected AI-generated explicit images on a single dark web forum in one month, 27\% of which were classified as illegal  under UK law \parencite{foundation2023a}. AI CSAM is not limited to the dark web—appearing also on social media, content-sharing sites, and subscription-based platforms \parencite{hingorani2023a, foundation2025a, thompson2024a}. 

In 2024, the IWF reported the emergence of AI CSAM \textit{videos} on monitored forums and a 22\% increase over six months in engagement with AI-specific forum threads \parencite{foundation2024a}. The same year, the National Center for Missing and Exploited Children [NCMEC] received five million more reports involving CSAM video compared to reports of CSAM images, underscoring video as a key vector for future harm \parencite{missing2025a}. Although generative video tools remain relatively nascent, creating friction for their misuse should be seen as a priority frontier for prevention.

In 2024, there was a 380\% increase in the number of reports made to the IWF containing actionable AI CSAM compared to the previous year \parencite{foundation2025a}. Alongside this dramatic growth, emerging evidence suggests a widening array of exploitative uses. One in ten minors surveyed reported knowing peers who had used generative AI to create explicit images of other children \parencite{thorn2024d}. Law enforcement professionals also flagged a rise in AI “nudify” tools used on minors, with 56\% of those surveyed having encountered such cases \parencite{bracket2024a}. In financial sexual extortion reports where tactics were identifiable, 11\% involved threats using fabricated sexual imagery of the child \parencite{thorn2024f}. This pace of change, along with AI CSAM’s scalability and realism pose challenges not only for detection and enforcement  \parencite{davy2024a, i2024b,thiel2023a} but also for understanding its broader societal and psychological implications \parencite{thorn2024e}. In this paper, we provide an overview of key concepts and technologies involved in the production of AI CSAM before examining known and potential harms associated with it.  

\subsection{Defining AI CSAM}
\label{subsec1}
AI CSAM refers to any sexually explicit visual depiction of a child, created or edited using techniques such as diffusion models, generative adversarial networks (GANs), or other AI-driven image and video synthesis technologies. Unlike CSAM that is created through the abuse of real children, AI CSAM is synthetically generated. However, its production can still victimize or revictimize real children, including through the use of CSAM in model training, through the use of AI in editing images of real children, or the creation of content that resembles identifiable minors. Some authors distinguish between \textit{AI-generated} CSAM and \textit{AI-manipulated} CSAM \parencite{krishna2024a} to capture the difference between fully synthetic versus partially synthetic AI-driven CSAM. Readers may be familiar with the term \textit{deepfake} which refers to partially synthetic content whereby camera-taken images or videos have been edited using AI tools \parencite{foundation2023a}. The AI CSAM videos initially observed in dark web forums in 2024 were mostly deepfakes or rudimentary fully synthetic videos \parencite{foundation2024a}. \textit{Nudifying} can be understood as a specific form of AI deepfaking, focused on synthetically removing clothing from material uploaded by the user. In practice, however, many nudifying tools extend beyond clothing removal, enabling the generation of sexual acts involving the uploaded individual, including video outputs created from still images \parencite{gibson2025a}. Nudifying and nudifying tools are frequently implicated in peer creation of AI CSAM as well as in explicit material created for sexual extortion of children and young people \parencite{thorn2025a}. 

\subsection{Situating AI CSAM within Theoretical Frameworks}
\label{subsec2}
Understanding the emergence and potential harms of AI CSAM requires situating it within broader theoretical frameworks of sexual offending. Seto’s Motivation-Facilitation Model (MFM) provides a well-established structure for understanding pathways to sexual offending, particularly differentiating between the motivational factors that drive an individual’s interest in CSAM and the facilitation and situational factors that enable offending behavior \parencite{seto2019a}. While sexual interest in children is a key motivational factor, it is not the only motivating factor for CSAM use (see Klein et al., \citeyear{klein2015a}, who report on the relationship between sex drive and CSAM use). In the MFM, offense patterns are shaped by facilitators, such as self-regulation problems and alcohol use, and situational factors, such as access to victims, and absence of supervision. Unchecked, AI CSAM may act as both a facilitation and situational factor within pathways to sexual offending, while also constituting an offense in its own right. As a facilitation mechanism, it offers material that reinforces sexual scripts and may lower inhibition through perceived normalization. As a situational factor, its accessibility, the anonymity afforded by downloadable and locally run models, and its customizability create low-friction opportunities to seek out, generate, or distribute abuse material. 

Lawless Space Theory (LST; Steel et al., \citeyear{steel2023a}) provides a complementary framework for understanding how offenders navigate digital environments and how these spaces shape offending behavior. LST suggests that online environments perceived as having weak governance—where anonymity, low enforcement risk, and limited accountability create conditions of perceived safety—increase the likelihood of sexually harmful behavior. The creation of tools capable of producing AI CSAM at scale presents an unprecedented expansion of lawless digital spaces, where offenders can produce illicit material with fewer logistical barriers and lower perceived risks—for example through their ability to download tools to generate novel material in a fully offline system, making it harder to detect offenses against children. The scalability and adaptability of AI tools also allow offenders to engage in hyper-personalized content creation, potentially reinforcing cognitive distortions and contributing to the normalization of child sexual exploitation.

Taken together, the MFM and LST frameworks underscore why AI CSAM is not merely an alternative form of illicit material but a fundamental shift in the mechanisms that facilitate child sexual exploitation. Rather than serving as a passive outlet, AI CSAM may function as both an accelerant for normalization and a reinforcement mechanism for deeper engagement in exploitative behaviors. These perspectives challenge arguments that AI CSAM could serve as a harm reduction tool, instead highlighting the ways in which such material extends existing risks and creates new avenues for victimization.

Understanding these facilitative and risk-enhancing properties of AI CSAM underscores why examining the technologies that enable it is critical. The following section outlines the specific mechanisms through which AI systems generate, modify, and distribute CSAM, lowering barriers to access and amplifying harm potential.

\section{Key Technologies}
\label{sec2}
For many, the technology that drives generative AI feels impenetrable or incomprehensible. This is true of frequent users of AI as well as those who have thus far avoided it. Science fiction writer Arthur C. Clarke famously posited that “any sufficiently advanced technology is indistinguishable from magic” (\citeyear{clarke1968a}; p. 255). The \textit{black box} between user prompt and machine output in generative AI may feel like magic to many people. This perception complicates discussions of AI CSAM, making it seem like an abstract ethical issue rather than a tangible harm. It can feel like a theoretical debate about whether AI-generated child abuse images and videos are just fictional depictions—a debate over whether fantasy is being criminalized or free expression restricted. However, understanding how AI CSAM is actually created allows for a more nuanced discussion of harm—one that goes beyond simply asking whether it is \textit{real} or \textit{not real}. Below, we explain some of the key technologies to inform later discussions around harm.  
\subsection{Diffusion Models}
\label{subsec3}
Diffusion models (such as Stable Diffusion, Midjourney, and DALL·E) are a class of generative models that transform random noise into coherent and detailed images or videos through a denoising process. These models can be combined with language models, allowing the resulting system to interpret user instructions—known as \textit{prompts}—that specify what kind of imagery to generate. First, the prompt is broken down into smaller pieces, called \textit{tokens}, which the model matches to visual concepts it has learned from billions of training images/videos and text descriptions of that imagery—such as shapes, colors, and textures associated with different labeled objects and scenes. As the model removes noise in each step, it shapes the imagery to better match the meaning of the prompt, gradually turning randomness into a clear and detailed visual media. An interactive explainer of this process can be found here: \url{https://poloclub.github.io/diffusion-explainer/} \parencite{lee2024a}. By learning patterns from large datasets, these models can synthesize entirely new images and videos that reflect the characteristics of the data they were trained on. Their ability to produce photorealistic imagery with fine-grained control has made them the dominant tools for generating synthetic content \parencite{yang2024a}, including AI CSAM \parencite{thorn2024a}.

In principle, prompting a mainstream diffusion model should not produce CSAM if that model is trained following best practices on training data curation. Companies applying established trust and safety policies should filter training data to detect and remove CSAM, such that the model cannot explicitly learn from direct representations of the material. Additionally, guardrails should be in place to prevent compositional generalization (whereby a model that has learned representations of both benign depictions of children and adult sexual material may be capable of synthesizing these concepts in harmful ways; Thorn \& All Tech is Human, \citeyear{thorn2024e}). However, this type of data curation is not comprehensively practiced across industry. The widely used training dataset LAION-5B—a dataset of 5 billion images, each paired with a descriptive caption—was found to contain links to CSAM \parencite{thiel2023b}. While the dataset was later updated to address this issue \parencite{l2024a}, earlier versions of Stable Diffusion were trained on data that included at least some illegal material. 

Fine-tuning tools such as DreamBooth, textual inversion, and LoRA (Low-Rank Adaptation) allow diffusion model users to customize models such that the outputs align with specific preferences. DreamBooth is a resource-intensive method of fine tuning that updates the entire model—which may have billions of parameters—to incorporate new styles, subjects, or domains. Textual inversion and LoRA are lightweight fine-tuning methods. Textual inversion teaches the model to associate a new token with a specific concept, allowing users to generate images of it. LoRA trains small additional layers to adjust the model’s behavior based on a small set of images (see Thiel et al., \citeyear{thiel2023a} for accessible explanations of these processes). While many uses of these methods are innocuous, they also enable bad actors to fine-tune models on depictions of specific children, or optimize the model to output imagery consisting of particular ages, poses, or explicit settings \parencite{thiel2023a}. 

While fine-tuning remains a key method of misuse, research indicates that illicit content can be generated without modifying the model itself. Instead, bad actors can exploit prompt engineering vulnerabilities, using carefully crafted textual inputs to bypass safety measures \parencite{he2024a}. These \textit{jailbreaking} techniques demonstrate how individuals can manipulate prompts to evade safety filters, significantly lowering the barrier for generating restricted content. However, in many cases, bad actors do not need to rely on jailbreaking at all. Loosely moderated diffusion model interfaces further reduce barriers to misuse by providing access to models built with minimal safeguards, and/or models with no filtering or content moderation on their outputs (e.g., Burgess, \citeyear{burgess2025a}). These platforms allow users to generate potentially harmful content without requiring technical expertise or modifications to the underlying AI model. 

Research demonstrates that fundamental training data curation can prevent undesirable capabilities and produce models that are more tamper-resistant, including to adversarial fine-tuning downstream \parencite{obrien2025a}. However, as noted earlier, this type of training data curation with respect to CSAM filtering is still not universally practiced, with many historical models that have not undergone this filtering still readily available \parencite{thorn2024a}.

The open-source and open-weight nature of many diffusion models exacerbates risks. Once released, these models can be freely modified and fine-tuned, making enforcement of safeguards extremely challenging. Even when developers implement safety features, users      can disable built-in content filters with minor code modifications, such as removing NSFW (not safe/suitable for work) detection mechanisms. Safety features that are more difficult to circumvent (such as ensuring the model is trained on properly curated data) can still be undone with adversarial fine-tuning or other methods \parencite{obrien2025a}. This means that, once a model is downloaded, there is no central authority capable of restricting its use. As a result, uncensored models—stripped of guardrails—are widely available online, ensuring continued access to unrestricted image generation \parencite{hawkins2025a}.

Another growing concern is the role of hybrid workflows, where diffusion models are used not only to create wholly synthetic imagery, but also to manipulate and enhance camera-taken imagery. Techniques such as inpainting and targeted image editing allow users to selectively alter parts of existing photographs—filling in masked areas or modifying specific features to produce realistic but deceptive content \parencite{mareen2024a}. This creates a dual risk. First, the blending of real and synthetic material can obscure the provenance of an image, undermining forensic investigations and impeding efforts to identify victims or sources. Second, it enables bad actors to layer abusive content onto innocuous imagery of real children, or to escalate the severity of existing abusive material. These workflows can combine diffusion-based inpainting with other generative techniques to increase realism and reduce detectability. 
\subsection{Generative Adversarial Networks}
\label{subsec4}
Generative Adversarial Networks (GANs) represents an older but still used generative AI technology. GANs consist of two neural networks: a generator, which creates synthetic data, and a discriminator, which evaluates the realism of the output. This adversarial process iteratively improves the quality of generated images, making GANs highly effective for photorealistic synthesis.

GANs remain widely used in hybrid workflows, particularly in cases where real and synthetic elements are combined \parencite{i2024b}. While diffusion models are increasingly favored for generating entirely synthetic CSAM, GAN-based techniques are still widely used for face-swapping or altering real images to produce explicit content.
\subsection{Emerging Technologies and Future Risks}
\label{subsec5}
While the major catalyst for concern and action to address AI CSAM has been the emergence and proliferation of text-to-image diffusion models, the broader generative AI ecosystem is evolving rapidly. Emerging systems are increasingly capable of generating full-motion video, 3D scenes, and highly customized or interactive multimodal content with minimal technical expertise. This increasing sophistication is likely to lower barriers to entry, reduce time and effort required to produce abuse material, and make it harder to distinguish real from synthetic media. Generative models are also becoming more directable and scalable, enabling the production of more targeted, prompt-specific content that could reflect specific individuals or scenarios. In parallel, agentic models capable of taking autonomous actions may introduce new risks, such as automating attempts to contact or groom children.
\section{Known and Potential Harms of AI CSAM}
\label{sec3}
The emergence of AI tools capable of producing synthetic CSAM has profound and far-reaching implications for victims, law enforcement, and broader societal attitudes toward child exploitation. We discussed above that links to CSAM have been identified in the training datasets used by diffusion models \parencite{thiel2023b}. This evidences harm and revictimization within the underlying architecture of some of these tools. In the following sections we explore seven ways in which outputs generated or manipulated by AI tools cause harm or have the potential to cause harm. The harms associated with AI CSAM are not limited to its direct impact on victims or those depicted. Several domains of harm emerge at the systemic level, including commercial and enforcement-related dynamics, that entrench abusive practices, impair protective responses, and increase long-term risks to children. To evidence harms we draw on industry and civil society reports, insights, and position papers (e.g., Thorn \& All Tech is Human \citeyear{thorn2024e}; Thiel et al., \citeyear{thiel2023a}); news reports; as well as sources making relevant arguments that predate the rise of diffusion models and the resulting wave of AI CSAM (e.g., Christensen et al., \citeyear{christensen2021a}).   

\begin{landscape}
\begin{table}[ht]
\centering
\caption{Summary of Known and Potential Harms Associated With AI CSAM}
{\small
\begin{tabular}{L{4.5cm} p{16cm}}
\toprule
\textbf{Category of Harm} & \textbf{Description} \\
\midrule
Depicting Real Children                                & Use of AI tools to generate explicit images that depict real children, including known abuse victims, minors whose images are in circulation, and children in the creator's immediate environment (online or offline). This results in ongoing (re)victimization, psychological distress, and exploitation, even for children who have been removed from abusive environments. \\
Coercion, Grooming, and Sexual Extortion               & Offenders exploit AI-generated explicit content to manipulate, desensitize, or blackmail children, increasing risks of grooming and coercion. Fabricated abusive images can be weaponized to construct false narratives, forcing victims into further exploitation. \\
Normalization and Desensitization                      & AI CSAM risks lowering psychological and social barriers to more extreme content. By normalizing child sexual exploitation, it may degrade users' moral and emotional inhibitions, reinforcing distorted beliefs and reducing perceived harm. \\
Gateway to Offending                                   & AI CSAM may serve as a behavioral bridge into offending through two mechanisms: (1) Escalation, where individuals progress from legal adult content to synthetic CSAM as tolerance builds toward more extreme material; and (2) Inhibition erosion, where individuals with a sexual interest in children, who might otherwise avoid offending, are drawn in by the \textit{perceived} safety, legality, or personalization of synthetic content. Both processes may be reinforced by online communities that normalize or encourage continued engagement. \\
Youth Access and Peer Exploitation                     & Adolescents are using AI tools to generate explicit images of peers, often without understanding the full consequences. This creates risks of coercion, abuse, and long-term psychological harm, while also implicating minors in digital sexual exploitation.  \\
Impaired Protection and Detection Capacity             & The sophistication of AI-generated content complicates law enforcement efforts to distinguish real from synthetic abuse images, complicating victim identification and increasing investigative burdens. AI-manipulated CSAM may obscure forensic details crucial for identifying at-risk victims.  \\
Incentivized Production and Profit-Driven Exploitation & AI CSAM is increasingly monetized with custom orders and AI models specifically designed for exploitation. The commercialization of synthetic CSAM fuels demand and entrenches exploitative economies, incentivizing further technological advancements for illicit purposes.   \\    \bottomrule
\end{tabular}}
\end{table}
\end{landscape} 

\subsection{Depicting Real Children}
\label{subsec6}
A direct concern is the use of AI CSAM to depict real children—either children who have been abused in the past to create CSAM or children who have not previously been victimized. Reports indicate that AI tools are being used to generate imagery featuring the likenesses of victims of past sexual abuse \parencite{foundation2023a, foundation2024a, thiel2023a}. In this way, even children who have been removed from abusive environments, or have their known CSAM imagery hashed for detection, continue to be victimized through new forms of synthetic exploitation \parencite{foundation2024a}. AI technology enables the creation of explicit material from non-explicit images including celebrity children or minors whose photos are shared innocently on social media \parencite{foundation2023a,foundation2024a} and children whose likeness is captured while they are in public spaces (e.g., Brewster, \citeyear{brewster2024a}). This material may be wholly novel (such as in the case of images created through the use of diffusion models fine-tuned on representations of a specific child) or reflect the use of “nudification" applications (see Gibson et al, \citeyear{gibson2025a}), deepfake tools, or other forms of AI-assisted editing of genuine images to create explicit images that depict specific children.

Professional organizations highlight the psychological risks associated with a child's likeness being used in AI CSAM, including the risk of humiliation, shame, anger, violation, and self-blame \parencite{pediatrics2025a}. Empirical research by \textcite{thorn2025a} further demonstrates that the overwhelming majority of teens and young adults (84\%) recognize deepfake nude images as causing tangible psychological, emotional, and reputational harm to those depicted. Victims report experiences of humiliation, violation, anxiety, and loss of control over their image, even when the material is entirely synthetic. 

As an emerging threat, it is not yet clear whether AI CSAM will be associated with distinct patterns of trauma among victims aware of their victimization. However, established psychological harms linked to the dissemination of CSAM—including symptoms of post-traumatic stress, shame, anxiety, and self-blame—are almost certain to persist when existing abusive material is multiplied, altered, or rendered more violent through the use of AI tools \parencite{chauvir2024a}. The creation of wholly new synthetic abuse, particularly when layered onto real prior victimization, introduces the possibility of qualitatively different trauma responses. Victims and survivors may face distortions of their own memories of abuse and experience a profound sense of helplessness in the face of abuse that not only resurfaces but mutates and proliferates.

\subsection{Coercion, Grooming, and Sexual Extortion}
\label{subsec7}
AI CSAM is not only passively consumed but also increasingly weaponized to facilitate new forms of coercion and exploitation. Reports highlight how manipulated or fully synthetic images are used along with other AI tools in grooming, with offenders leveraging fabricated explicit material to desensitize children, threaten exposure, or carry out further exploitation \parencite{coalition2025a,qoria2024a}. A growing concern is the rise of sexual extortion cases involving AI-generated imagery \parencite{milmo2025a}, where synthetic sexual images are used to manipulate or coerce minors into producing real sexually explicit material, paying money, or providing access to sensitive information in lieu of payment \parencite{aronashvili2024sextortion, investigation2023a, thorn2024c}. These attacks may involve purely synthetic material, real material, or a combination of both—such as when AI-generated or manipulated content is used to coerce victims into creating new, real imagery that is subsequently exploited. They may also include only threats to create AI manipulated content. 

Online sexual extortion offenders are characterized by a variety of motivations, including individuals with a sexual interest in children as well as financially-motivated individuals who may engage in the targeting of young people as part of a wider range of cybercrime and scam behaviors \parencite{o2022a}. This blurring of cybercrime types risks the introduction of a wider range of AI-facilitated techniques—such as voice cloning to enhance credibility—with which to perpetrate sexual exploitation at scale \parencite{investigation2024a,raffile2024a}. While the role of AI in sexual extortion schemes is still evolving, the psychological impacts of sexual extortion on minors are already well-documented. Victims commonly experience trauma symptoms including shame, helplessness, anxiety, suicidal ideation, and difficulties trusting others \parencite{ray2024a,wolbers2025a}. Sexual extortion has been linked to self-harm and suicide in a number of cases \parencite{ray2024a}. 

\subsection{Normalization and Desensitization}
\label{subsec9}
The widespread availability of AI CSAM also risks desensitization and normalization of child sexual exploitation. There is increasing concern that engagement with synthetic material lowers the psychological threshold for individuals to seek out more extreme content, increasing the risk of transitioning to real-world abuse \parencite{thiel2023a,parti2024a}. Research on CSAM offenders suggests that those who consume abusive material have belief systems that minimize harm rather than explicitly endorse abuse, allowing them to justify continued engagement \parencite{bartels2016a}. AI CSAM may reinforce a similar mechanism, providing users with a justification for ongoing use under the belief that synthetic content is fundamentally different from camera-taken CSAM, even as it maintains the same underlying exploitative themes. This misperception may be widespread: a UK survey found that 40\% of adults were unsure whether AI CSAM was legal or believed it to be legal \parencite{foundation2024b}. Such uncertainty around legality could further lower users’ perceived accountability and reduce the perceived harm of engaging with synthetic abusive material.

While some population-level studies have suggested that increased access to pornography including CSAM may correlate with lower rates of sexual offending including child abuse (e.g., Diamond et al., \citeyear{Diamond2011}), interpretation of these data are challenging given the use of natural—rather than controlled—experiments for evidence. More recent work indicates that, among users of CSAM, higher frequency of use and exposure to more extreme material are associated with increased likelihood of seeking direct contact with children, particularly in online spaces \parencite{Insoll2022}. The findings of \textcite{Insoll2022} also suggested that a substantial portion of their sample \textit{feared} that their use of CSAM or other extreme materials would lead to an escalation in sexual behavior. 

 A feature of AI CSAM is that it not only facilitates access to existing abusive material but also expands a potential range of harm by enabling the creation of abuse scenarios that would otherwise be difficult to access in non-AI CSAM. Reports indicate that AI tools are being used to generate hyper-violent, sadistic, or otherwise extreme material that may not exist, or may not exist in similar volumes, in real-world CSAM collections \parencite{foundation2024a}. This is particularly concerning because AI CSAM is no longer constrained by what has previously been documented; it allows offenders to create entirely new forms of abuse content that align with their "wildest fantasies" as described in offender discussions on dark web forums \parencite{foundation2024a}. Examples shared by \textcite{missing2024a} demonstrate real prompts seeking to generate material that would correspond to the highest levels of severity on the COPINE rating scale of sexual abuse imagery \parencite{taylor2001a}. The ability to construct such content on demand raises concerns about not just normalization but escalation, where access to more extreme synthetic content could further degrade psychological inhibitions against real-world abuse.

\subsection{Gateway to Offending}
\label{subsec10}
AI CSAM risks enhancing existing or opening new pathways into offending, either by facilitating the escalation of pornography consumption toward more extreme material or by reducing protective barriers that would otherwise prevent engagement with CSAM by people with a longstanding sexual interest in children. Research suggests that some individuals who engage in compulsive pornography use or hypersexual behavior experience tolerance effects, where they seek out increasingly novel or taboo material \parencite{seto2019a}. While not all individuals progress to illegal content, evidence from CSAM offender studies indicates that prolonged engagement with pornography can, for some, lead to the normalization of more extreme content, including CSAM \parencite{knack2020a}. AI CSAM may accelerate this process by offering a form of highly personalized and readily available material that removes previous safeguards against escalation. Loosely moderated online interfaces and offline models with minimal restrictions blur the boundaries between adult-oriented and child sexual content, creating new routes through which individuals may shift from legal to illegal material.

For individuals with a sexual interest in children who have strong protective factors—such as moral beliefs or fear of legal consequences—AI CSAM introduces a perceived "safe" outlet that could erode these inhibitions over time. Research on CSAM offenders suggests that many justify their behavior not by overtly endorsing abuse but by minimizing harm \parencite{bartels2016a}. While research on AI CSAM is still emerging, prior studies of online CSAM offenders indicate that engagement in digital sexual exploitation materials can facilitate normalization of offending behaviors. \textcite{elliott2009a} describe how online environments provide access to social validation, justification, and reinforcement for CSAM use, which can diminish psychological and social barriers to further offending. AI CSAM exists within these same digital spaces, meaning those engaging with it may also be exposed to content and communities that normalize further offending behaviors. A report by the \textcite{foundation2024a} described “some users discuss sharing AI-generated images with non-perpetrators as an intended ‘gateway’ to real CSAM” (p. 35). 

\subsection{Youth Access and Peer Exploitation}
\label{subsec11}
A further concern is the accessibility of certain AI tools to adolescents, lowering the barriers to producing explicit images of peers. Reports indicate that young people are using AI software to create non-consensual explicit images of classmates, often without full comprehension of the consequences \parencite{hale2025a,laird2024a,thorn2024d,thorn2025a,centre2023a}. Recent research from the U.S. further confirms this trend: a small but non-trivial number of adolescents report using AI tools found via app stores, social media, or general web searches to generate deepfake nude images of peers, motivated by arousal, curiosity, peer pressure, or as a way to enact revenge or to bully \parencite{thorn2025a}. Once created, these images can escape the original peer context—through sharing, leaks, or online circulation—leading to repeated victimization and raising concerns about broader dissemination and potential secondary exploitation.

These technologies also add a new layer of perpetration risk by transforming what might otherwise remain private or normative sexual thoughts—such as fantasizing about peers—into tangible, distributable, and harmful content. In doing so, they collapse the boundary between internal fantasy and external action, exposing adolescents to legal jeopardy and social consequences in ways that earlier generations did not face. Overall, the ease of generating explicit content within adolescent peer groups raises urgent concerns about social and psychological harm, legal jeopardy, reputational damage, and the potential for escalation to coercion, blackmail, or further abuse.

\subsection{Impaired Protection and Detection Capacity}
\label{subsec12}
The increasing sophistication of AI tools also creates significant challenges for law enforcement in distinguishing between real and synthetic material \parencite{foundation2024a, foundation2023a}. As AI-generated images become more photorealistic, identifying whether an image depicts an actual child in need of protection becomes more difficult \parencite{crawford2023a}. The time and resources required to verify whether material is real or synthetic delays responses to cases involving children experiencing ongoing or imminent abuse, diverting investigative focus, and placing additional strain on forensic units \parencite{foundation2023a,parti2024a,thiel2023a}. The challenge posed by AI capabilities for law enforcement is not limited to identifying where AI has been used in the creation or addition of sexual content to images and video, but also the identification of AI-manipulated CSAM which may erase or obscure key forensic details that investigators rely on to identify real children and their locations \parencite{thiel2023a}. These investigative barriers and delays directly undermine the protection of children, allowing some real-world abuse to continue undetected while synthetic content diverts critical forensic resources.

The strain on the child protection ecosystem is compounded by the fact that CSAM reports submitted to clearing houses such as NCMEC do not always include labels to indicate whether an image was AI-generated or AI-modified. While some tech platforms include this information under certain conditions—such as when confidence is high in their classification of AI use, or when the material was verifiably produced by their own AI services—others may not apply such processes consistently, or at all. This may stem from workload challenges, lack of available metadata or signal, or from an understandable concern over incorrectly labelling something as AI, particularly where the image depiction or context might indicate a child at imminent risk. The overall result, however, may be a missed opportunity to more effectively triage and prioritize cases, further overwhelming law enforcement and child protection organizations and thus increasing the difficulty in identifying at-risk children. 

\subsection{Incentivized Production and Profit-Driven Exploitation}
\label{subsec13}
Another major risk is the commercialization of AI CSAM, a trend increasingly documented by journalists and child protection organizations \parencite{crawford2023a,foundation2023a, koltai2024opendream}. The IWF has documented cases where AI-generated abuse material is being monetized, either through the sale of pre-made images or through custom orders that cater to specific exploitative preferences \parencite{foundation2023a}. In some cases, AI models trained explicitly for CSAM generation have been distributed within underground networks, raising concerns that these tools are fueling demand for real-world abuse material \parencite{foundation2023a,foundation2024a}. Law enforcement responses have begun to reflect these concerns, with recent prosecutions involving individuals who used AI tools to create and sell synthetic abuse images \parencite{service2024a}, and coordinated international operations targeting creators and distributors of AI CSAM \parencite{europol2025a}. The presence of a commercial market for synthetic CSAM further entrenches exploitative economies and incentivizes the continued development of more advanced AI tools for illicit purposes, as has already been observed in cases of monetized synthetic CSAM shared across both underground and mainstream platforms \parencite{crawford2023a,foundation2023a}. These dynamics contribute to the persistence and proliferation of child sexual exploitation material, expanding both its reach and its psychological toll on those targeted or depicted.

\section{The AI CSAM \textit{as harmless} counterargument}
\label{sec4}
While there has been a major global focus on the threats associated with generative AI, there has also been sustained attention on how AI tools can help combat online and offline sexual harm \parencite{unicri2023_ai_children, steel2024a}. It is therefore unsurprising that individuals have considered whether AI CSAM has potential use as an alternative or substitute for other CSAM as a way of managing urges to commit other sexual offenses. While unequivocal advocacy for this position is rare in peer-reviewed literature, partial support or consideration of the argument appear in science-focused popular media sources (e.g., Bernstein, \citeyear{bernstein2023a}; Maier, \citeyear{maier2022a}). The argument is acknowledged—albeit critically—in sources that aim to refute it, such as \textcite{sheepshanks2024a} and \textcite{thiel2023a}. It is also explored in the context of pre-diffusion model virtual CSAM, where researchers examine both the harm-reduction hypothesis and the risks of reinforcement or escalation \parencite{christensen2021a}. Support for versions of the argument are also voiced—and debated—in some web forums, including those aimed at individuals with a sexual interest in children (e.g., VirPed) and more general online spaces (e.g., Reddit). Anecdotally, this argument has also been a source of friction within tech industry organizations, where AI-generated sexual material involving fictional minors is sometimes viewed as a lower priority for trust and safety efforts due to perceptions of its harm relative to other CSAM, or due to concerns about free expression.

In relevant work, Moen and Sterri (\citeyear{moen2018a}, see also Moen, \citeyear{moen2015a}, and Sterri \& Earp, \citeyear{sterri2021a}), engaged in a philosophical exploration of whether certain forms of sexual material that do not directly involve real children—such as child sex robots, fictional depictions of child abuse, or computer-generated CSAM—could serve as an alternative that prevents real-world harm. While their arguments predate the recent advancements in generative AI, they have been echoed in some contemporary discussions around AI CSAM. The core arguments suggest that: (1) where no real child is involved, the production and use of such material may not be inherently harmful; and (2) in some cases, access to non-contact outlets could serve a harm reduction function by preventing real-world offenses.

These arguments warrant careful consideration, particularly in light of our contention that AI CSAM does not operate in isolation but actively contributes to a broader ecosystem of harm. As established earlier, there is CSAM in the provenance of some widely used image generation models through its inclusion in their training \parencite{thiel2023b}. Rather than existing as a separate category of content to CSAM, AI CSAM implicates real children, particularly when it is used to fabricate explicit images of identifiable minors \parencite{foundation2023a,foundation2024a,mccrindle2024a,thiel2023a}. The customizable nature of AI CSAM means that the content can be rendered more violent, sadistic, and otherwise extreme than any underpinning CSAM used in its creation. This customizability also lends itself to the creation of material that is more emotionally engaging for users, for example through the use of chatbots and other multimodal AI tools. This capacity reinforces concerns that synthetic content contributes to escalating patterns of consumption. \textcite{christensen2021a} argued that users of virtual CSAM may experience fantasy escalation, whereby tolerance builds over time and leads to the pursuit of more extreme or real-world material. Studies of CSAM offenders indicate that prolonged engagement with abusive material can be associated with escalation to more extreme content and, in some cases, seeking contact with children \parencite{Insoll2022}. Reports from the \textcite{foundation2024a} further suggest that some offenders view AI CSAM as a gateway to CSAM, raising concerns that it may erode psychological and moral barriers to further harm.

The harm reduction argument advanced in earlier discussions also assumes that engagement with synthetic CSAM operates in a way that parallels harm reduction approaches in other domains, such as regulated access to controlled substances. However, this analogy breaks down when applied to the psychological and social mechanisms underlying child sexual exploitation. Unlike controlled substances, which may be provided in measured doses under supervision, AI CSAM is highly scalable, easily distributed, and—crucially—used in private, unregulated contexts that may enable rather than constrain harmful behavior. As \textcite{christensen2021a} argue, the perceived anonymity of virtual CSAM use may reduce users’ inhibitions and foster distorted rationalizations, such as the belief that ‘no one is harmed.’ While evidence of a direct link between harm minimizing beliefs and offending risk is scant, they are theorized to lower psychological barriers and sustain engagement. Moreover, online communities that tolerate or endorse synthetic CSAM can provide social reinforcement, further entrenching these beliefs and reducing the likelihood that individuals will seek help or perceive their behavior as problematic. 

These dynamics challenge the view that synthetic CSAM can serve as a controlled or protective outlet. They also highlight how real-world developments have outpaced earlier philosophical arguments. \textcite{moen2018a}, writing before the advent of diffusion-based generative AI, explored whether fictional or computer-generated sexual material could serve as a non-harmful outlet. However, current generative AI technology allows for the creation of abuse scenarios that are more violent, sadistic, or emotionally resonant than those previously imaginable \parencite{foundation2024a}. This shift introduces new risks of desensitization and behavioral escalation, not simply by enabling access to abusive material but by expanding the very parameters of what that material can portray.

Furthermore, while prior arguments for synthetic CSAM have largely focused on adult consumers, the increasing accessibility of generative tools to adolescents introduces a different category of harm. Evidence of non-consensual peer use—where young people use AI to create explicit images of classmates \parencite{hale2025a, thorn2025a, centre2023a}—further undermines the idea that synthetic CSAM can be ethically contained within a harm reduction framework. These developments extend the spectrum of risk well beyond the scenarios envisioned in early philosophical debates.

Ultimately, while \textcite{moen2018a} frame their discussions within an exploration of minimizing harm, and while AI CSAM specific discussions mirror these, the emerging reality of AI CSAM suggests that these theoretical arguments do not translate into practice. The harms outlined earlier in this paper demonstrate that synthetic CSAM is not a neutral or lesser alternative to CSAM; rather, it introduces new forms of victimization, potentially contributes to behavioral escalation, and creates significant enforcement challenges. These risks appear to substantially outweigh any speculative benefit that AI-generated content might provide. This underlines the importance of challenging harmlessness narratives where they appear to be the default view. 

This argument is not, however, incompatible with a view in which there may be individuals for whom a carefully controlled form of synthetic material—distinct from AI CSAM—could have a role in helping them live offense-free lives. Any such exploration would need to remain within clear legal and ethical boundaries, including ensuring that material cannot depict real children or be built from representations of them. Whether such resources could serve a legitimate therapeutic or preventative function—and if so, whether that function is sexual or grounded more in relational needs—requires careful ethical scrutiny. Research in this area would need to uphold strict standards of evidence and clearly delineate the boundaries between harm reduction practices and those that risk direct or indirect harm. The claim that AI CSAM is harmless or less harmful remains extraordinary—and, as such, demands extraordinary evidence. 

\section{Conclusion}
\label{sec5}
Discussions around AI CSAM are often shaped by an appeal to harmlessness. Compared to CSAM, which involves direct exploitation, AI-generated material may appear—at first glance—to exist at a distance from harm. This assumption may influence public discourse, industry priorities, and, in some cases, regulatory inaction (e.g., Schurig \& Granjeia, \citeyear{schurig2024a}). Yet, as this paper has demonstrated, the notion that AI CSAM is inherently less harmful rests on a fundamental misunderstanding: both of the ways harm manifests and of how this material is actually created.

One source of this misunderstanding is naivety about the breadth of harm associated with AI CSAM. The production of synthetic material does not occur in a vacuum; it is built upon datasets containing existing CSAM and/or images of real children. AI CSAM is not simply the output of a neutral algorithm generating fictional depictions—it is shaped by past victimization, fine-tuned using real abuse material, and sometimes designed to resemble known victims. Beyond its production, AI CSAM is embedded within ecosystems of exploitation, where it reinforces cognitive distortions, facilitates grooming and extortion, and, in some cases, serves as a bridge toward contact offending.

A second form of naivety is a tendency to evaluate harm in relative rather than absolute terms. Because wholly AI-generated CSAM does not involve the immediate suffering of a child during its creation, it is often perceived as a lesser issue when compared to direct physical abuse. This relative framing creates a false dichotomy—one that ignores the ways AI CSAM fuels broader cycles of exploitation and desensitization. Just as non-contact sexual offenses (e.g., CSAM possession) are not "harmless" simply because they lack physical interaction, AI CSAM cannot be dismissed on the grounds that it does not involve direct abuse in the moment of its generation. Harm does not exist on a single dimension, and an exclusive emphasis on direct victimization risks obscuring the many other ways harm unfolds.

The challenge, then, is not only to recognize AI CSAM as harmful but to articulate that harm in ways that cut through these intuitive appeals to harmlessness. The synthesis presented in this paper serves as a resource for stakeholders—whether in law enforcement, policy, tech industry regulation, or child protection—who need to justify action in their respective domains. This justification does not hinge on speculative concerns about what AI CSAM \textit{might} lead to in the future; rather, it is grounded in the well-established mechanisms by which sexual exploitation material contributes to risk. AI CSAM is not a neutral technological artifact—it is an active facilitator of harm.

Recognizing this is essential for moving beyond the inertia that often accompanies emerging forms of abuse. Inaction is not always the product of disagreement; it is sometimes the result of a failure to articulate harm clearly enough that it demands a response. By confronting the narratives that have allowed AI CSAM to be seen as a lesser concern, this paper provides the foundation for that response.

\section*{CRediT authorship contribution statement}
Caoilte Ó Ciardha: Conceptualization,  Funding Acquisition, Writing - Original Draft Preparation, Writing - Review \& Editing; John Buckley: Conceptualization, Writing - Review \& Editing; Rebecca S. Portnoff: Writing - Review \& Editing.

\section*{Declaration of generative AI and AI-assisted technologies in the writing process}
During the preparation of this work the authors used ChatGPT (versions 4 and 5) and NotebookLM in order to help synthesize literature, simplify arguments and concepts, refine writing, brainstorm, and challenge conclusions. After using these tools, the authors reviewed and edited the content as needed and take full responsibility for the content of the published article.

\section*{Funding}
This work was supported by the Tech Coalition Safe Online Research Fund [23-EVAC-0015.2-University of Kent]. The funder had no role in the composition or drafting of this manuscript.

\section*{Declaration of Competing Interest}
John Buckley is a former Head of Child Safety at Google, a company involved in the development and commercial deployment of generative AI tools. His authorship of this paper is in a personal capacity and is not intended to reflect the views or process of Google in any way. Dr. Rebecca Portnoff is employed by Thorn, a child safety nonprofit that collaborates with and provides services to several technology companies, some of which are involved in the development or deployment of generative AI tools. Her contribution to this paper is made in a personal capacity and does not necessarily reflect the views or positions of Thorn or its partners. The authors otherwise declare no known competing financial interests or personal relationships that could have appeared to influence the work reported in this paper.

\printbibliography
\end{document}